\documentclass[11pt]{article}
\usepackage{amssymb}

\usepackage{graphicx}
\usepackage{amsmath}
\usepackage{makeidx}
\usepackage{indentfirst}

\newcounter{resultnum}[section]
\newtheorem{conclusion}{Conclusion}[section]

\newcounter{conclusionnum}[section]

\newcounter{conditionnum}[section]

\newcounter{conjecturenum}[section]

\newcounter{examplenum}[section]

\newcounter{exercisenum}[section]
\newtheorem{lemma}{Lemma}[section]

\newcounter{lemmanum}[section]

\newcounter{notationnum}[section]
\newtheorem{theorem}{Theorem}[section]

\newcounter{theoremnum}[section]
\newtheorem{definition}{Definition}[section]

\newcounter{definitionnum}[section]

\newcounter{corollarynum}[section]
\newtheorem{remark}{Remark}[section]

\newcounter{remarknum}[section]
\newtheorem{proposition}{Proposition}[section]

\newcounter{propositionnum}[section]

\newcounter{acknowledgementnum}[section]

\newcounter{algorithmnum}[section]

\newcounter{axiomnum}[section]

\newcounter{casenum}[section]

\newcounter{claimnum}[section]

\newcounter{summarynum}[section]

\newcounter{problemnum}[section]

\begin{document}

\title{Fractional  Almost K\"{a}hler -- Lagrange Geometry}
\date{September 26, 2010}
\author{\textbf{Dumitru Baleanu}\thanks{%
On leave of absence from Institute of Space Sciences, P. O. Box, MG-23, R
76900, Magurele--Bucharest, Romania; dumitru@cancaya.edu.tr,
baleanu@venus.nipne.ro} \\
\textsl{\small Department of Mathematics and Computer Sciences,} \\
\textsl{\small \c Cankaya University, 06530, Ankara, Turkey } \\
\and 
\textbf{Sergiu I. Vacaru} \thanks{%
sergiu.vacaru@uaic.ro, Sergiu.Vacaru@gmail.com} \and \textsl{\small Science
Department, University "Al. I. Cuza" Ia\c si, } \\
\textsl{\small 54, Lascar Catargi street, Ia\c si, Romania, 700107 } }
\maketitle

\begin{abstract}
The goal of this paper is to encode equivalently the fractional Lagrange
dynamics as a nonholonomic almost K\"{a}hler geometry. We use the fractional
Caputo derivative generalized for nontrivial nonlinear connections
(N--connections) originally introduced in Finsler geometry, with further
developments in Lagrange and Hamilton geometry. For fundamental geometric objects induced
canonically by regular Lagrange functions, we construct compatible almost
symplectic forms and linear connections completely determined by a "prime"
Lagrange (in particular, Finsler) generating function. We emphasize the
importance of such constructions for deformation quantization of fractional
Lagrange geometries and applications in modern physics.

\vskip0.1cm


\vskip3pt


\end{abstract}



\section{Introduction}

The fractional calculus and geometric mechanics are in areas of many
important researches in modern mathematics and applications. During the last
 years, various developments of early approaches to fractional calculus \cite%
{oldham,nishimoto,riewe97,carp,podlubny,hilfer,kilbas,
west,mainardi,momani,enrico,magin,lorenzo,agrawal,machado,
agrawal1,gorenflo,chen,orti}
were reported in various branches of science and engineering. The
fractional calculus is gaining importance due to its superior
results in describing the dynamics of complex systems
\cite{hilfer,magin,west}. One important branch of fractional
calculus is devoted to the fractional variational principles and
their applications in physics and control theory
\cite{bal06,balmus05,taras06,taraszasl,baltr}. One of the open
problems in the area of fractional calculus is to find an
appropriate geometrization of the fractional Lagrange mechanics.

In our paper we provide a geometrization of the fractional
Lagrange mechanics using methods of Lagrange--Finsler and (almost)
K\"{a}hler
geometry following the ideas and constructions from two partner works \cite%
{vrfrf,vrfrg} which initiated a new direction of fractional differential
geometries and related mathematical relativity/gravity theory and nonholonomic Ricci flow evolution.
Quantization of such theories is an interesting and challenging problem
 connected to fundamental issues in modern quantum gravity and fundamental
interaction/ evolution theories.

There were published different papers related to quantum fractional theories
(see, for instance, Refs. \cite{lask1,lask2,nab04}). Nevertheless, there is
not a general/systematic approach to quantization of fractional
field/evolution theories. The aim of our third partner work \cite{vdefqfrt}
is to prove that the strategy used in Refs. \cite{vfedq1,vfedq2,vfedq3,vfed4}
allows us to quantize also a class of fractional geometries following
Fedosov method \cite{fed1,fed2}. To perform such a program is necessary to
elaborate (in this paper) a general fractional version of almost K\"{a}hler
geometry and then to apply (in the next paper \cite{vdefqfrt}) the
Karabegov--Schlichenmaier deformation quantization scheme \cite{karabeg1}.
For the canonical models (with the so--called canonical Cartan--Finsler
distinguished connections \cite{cartan}) of Finsler and Lagrange geometries %
\cite{ma1}, there is a standard scheme of encoding such spaces into almost K%
\"{a}hler geometries (the result is due to Matsumoto \cite{matsumoto}, for
Finsler geometry, and then extended to Lagrange spaces \cite{opr1}; see
further developments in \cite{vbrane,anastv}).

In brief, a geometrization of a physical theory in terms of an almost K\"{a}hler geometry means a "natural" encoding of fundamental physical objects into terms of basic geometric objects of an almost symplectic geometry (with certain canonical one- and two--forms and connections, almost complex structure etc induced in a unique form, for instance, by a Lagrange or Hamilton function). In general, physical theories are with nonholonomic (equivalently, anholonomic, or non--integrable) structures which does not allow us to transform realistic physical models into (complex) symplect/K\"{a}hler geometries. Nevertheless, the bulk of physical theories can be modelled as almost symplectic geometries for which there were elaborated advanced mathematical approaches and formalism  of geometric/deformation quantization, spinor and noncommutative generalizations, unification schemes etc. Our main idea is to show that certain versions of fractional calculus (based on Caputo derivative), and related physical models, admit almost K\"{a}hler type fractional geometric encoding which provides an unification scheme with integer and non--integer derivatives for classical and quantum physical theories.

In our papers \cite{vrfrf,vrfrg,vdefqfrt}, we work with the Caputo
fractional derivative. The advantage of using this fractional derivative
(resulting in zero values for actions on constants) is that it allows us to
develop a fractional differential geometry which is quite similar to
standard (integer dimensions) models and to use standard boundary conditions
of the calculus of variations. There is also a  series expansion
formalism for the fractional calculus and fractional differential equations %
\cite{lrz} allowing us to compute fractional modifications of formulas
in quite simple forms. This explains why such constructions are popular in engineering
and physics.\footnote{%
 Perhaps, it is not possible
to introduce a generally accepted definition of fractional derivative. For
constructions with various types of distributions, in different theories of
physics, economics, biology etc, when the geometric picture is less
important, it may be more convenient to use the Riemann--Liouville (RL)
fractional derivative.}

The plan of this paper is as follows:\ In section \ref{s2}, we outline the
necessary formulas on fractional Caputo calculus and related geometry of
tangent bundles. A brief formulation of fractional Lagrange--Finsler
geometry is presented in section \ref{s3}. The main results on almost K\"{a}%
hler models of fractional Lagrange spaces are provided in section \ref{s4}.
Finally, conclusions are presented in section \ref{s5}.

\section{Preliminaries: Fractional Caputo Calculus on Tangent Bundles}

\label{s2} We follow the conventions established in Refs. 
\cite{vrfrf,vrfrg}.\footnote{All left, "up" and/or "low" indices will be considered as labels for some geometric objects but the right ones as typical abstract or coordinate indices which may rung values for corresponding space/time dimensions. Boldface letters will be used for spaces endowed with nonlinear connection structure.} In our approach, we try to elaborate fractional geometric models with
nonholonomic distributions when the constructions are most closed to ''integer'' calculus.

Let $f(x)$ \ be a derivable function $f:[\ _{1}x,\ _{2}x]\rightarrow \mathbb{%
R},$ for $\mathbb{R\ni }\alpha >0,$ and $\partial _{x}=\partial /\partial x$
be the usual partial derivative. By definition, the left Riemann--Liouville
(RL) derivative is
\begin{equation*}
\ \ _{\ _{1}x}\overset{\alpha }{\partial }_{x}f(x):=\frac{1}{\Gamma
(s-\alpha )}\left( \frac{\partial }{\partial x}\right) ^{s}\int\limits_{\ \
_{1}x}^{x}(x-\ x^{\prime })^{s-\alpha -1}f(x^{\prime })dx^{\prime },
\end{equation*}%
where $\Gamma $ is the Euler's gamma function. It is also considered the
left fractional Riemann--Liouville derivative of order $\alpha ,$ where $\ s-1<\alpha
<s,$ with respect to coordinate $x$ is $\ \overset{\alpha }{\partial }%
_{x}f(x):=\lim_{_{1}x\rightarrow -\infty }\ \ _{\ _{1}x}\overset{\alpha }{%
\partial }_{x}f(x).$

Similarly, the right RL derivative is introduced by
\begin{equation*}
\ _{\ x}\overset{\alpha }{\partial }_{\ _{2}x}f(x):=\frac{1}{\Gamma
(s-\alpha )}\left( -\frac{\partial }{\partial x}\right)
^{s}\int\limits_{x}^{\ _{2}x}(x^{\prime }-x)^{s-\alpha -1}f(x^{\prime
})dx^{\prime }\ ,
\end{equation*}%
when the right fractional Liouville derivative is computed $\ _{\ x}\overset{%
\alpha }{\partial }f(x^{k}):=\lim_{_{2}x\rightarrow \infty }\ \ _{\ x}%
\overset{\alpha }{\partial }_{\ _{2}x}f(x).$

Integro--differential constructions based only on RL derivatives seem to be
very cumbersome for the purpose to elaborate (fractional) differential
geometric models and has a number of properties which are very different
from similar ones for the integer calculus. This problem is a consequence of
the fact that the fractional RL derivative of a constant $C$ is not zero
but, for instance, $\ _{\ _{1}x}\overset{\alpha }{\partial }_{x}C=C\frac{%
(x-\ _{1}x)^{-\alpha }}{\Gamma (1-\alpha )}.$ We shall use a different
type of fractional derivative.

The fractional left, respectively, right Caputo derivatives are defined
\begin{eqnarray}
&&\ _{\ _{1}x}\overset{\alpha }{\underline{\partial }}_{x}f(x):=\frac{1}{%
\Gamma (s-\alpha )}\int\limits_{\ \ _{1}x}^{x}(x-\ x^{\prime })^{s-\alpha
-1}\left( \frac{\partial }{\partial x^{\prime }}\right) ^{s}f(x^{\prime
})dx^{\prime };  \label{lfcd} \\
&&\ _{\ x}\overset{\alpha }{\underline{\partial }}_{\ _{2}x}f(x):=\frac{1}{%
\Gamma (s-\alpha )}\int\limits_{x}^{\ _{2}x}(x^{\prime }-x)^{s-\alpha
-1}\left( -\frac{\partial }{\partial x^{\prime }}\right) ^{s}f(x^{\prime
})dx^{\prime }\ .  \notag
\end{eqnarray}%
In \ the above formulas, we underline the partial derivative symbol, $%
\underline{\partial },$ in order to distinguish the Caputo differential
operators from the RL ones with usual $\partial .$ For a constant $C,$ for
instance, $_{\ _{1}x}\overset{\alpha }{\underline{\partial }}_{x}C=0.$

The fractional absolute differential $\overset{\alpha }{d}$ is written in
the form
\begin{equation*}
\overset{\alpha }{d}:=(dx^{j})^{\alpha }\ \ _{\ 0}\overset{\alpha }{%
\underline{\partial }}_{j},\mbox{ where }\ \overset{\alpha }{d}%
x^{j}=(dx^{j})^{\alpha }\frac{(x^{j})^{1-\alpha }}{\Gamma (2-\alpha )},
\end{equation*}%
where we consider $\ _{1}x^{i}=0.$ For the ''integer'' calculus, we use as
local coordinate co-bases/--frames the differentials $dx^{j}=(dx^{j})^{%
\alpha =1}.$ If $0<\alpha <1,$ we have $dx=(dx)^{1-\alpha }(dx)^{\alpha }.$
The ''fractional'' symbol $(dx^{j})^{\alpha },$ related to $\overset{\alpha }%
{d}x^{j},$ is used instead of usual ''integer'' differentials (dual
coordinate bases) $dx^{i}$ for elaborating a co--vector/differential form
calculus. We consider the exterior fractional differential $\overset{\alpha }%
{d}=\sum\limits_{j=1}^{n}\Gamma (2-\alpha )(x^{j})^{\alpha -1}\ \overset{%
\alpha }{d}x^{j}\ \ _{\ 0}\overset{\alpha }{\underline{\partial }}_{j}$ and
write the exact fractional differential 0--form as a fractional differential
of the function $\ _{\ _{1}x}\overset{\alpha }{d}_{x}f(x):=(dx)^{\alpha }\ \
_{\ _{1}x}\overset{\alpha }{\underline{\partial }}_{x^{\prime }}f(x^{\prime
}).$ The formula for the fractional exterior derivative is
\begin{equation}
\ \ _{\ _{1}x}\overset{\alpha }{d}_{x}:=(dx^{i})^{\alpha }\ \ _{\ _{1}x}%
\overset{\alpha }{\underline{\partial }}_{i}.  \label{feder}
\end{equation}

For a fractional differential 1--form $\ \overset{\alpha }{\omega }$ with
coefficients $\{\omega _{i}(x^{k})\}$ we may write
\begin{equation}
\overset{\alpha }{\omega }=(dx^{i})^{\alpha }\ \omega _{i}(x^{k})
\label{fr1f}
\end{equation}%
when the exterior fractional derivatives of such a fractional 1--form $\
\overset{\alpha }{\omega }$ \ results a fractional 2--form,
$\ _{\ _{1}x}\overset{\alpha }{d}_{x}(\ \overset{\alpha }{\omega }%
)=(dx^{i})^{\alpha }\wedge (dx^{j})^{\alpha }\ _{\ _{1}x}\overset{\alpha }{%
\underline{\partial }}_{j}\ \omega _{i}(x^{k}).$ The fractional exterior
derivative (\ref{feder}) \ can be also written in the form
\begin{equation*}
\ \ _{\ _{1}x}\overset{\alpha }{d}_{x}:=\frac{1}{\Gamma (\alpha +1)}\ \ _{\
_{1}x}\overset{\alpha }{d}_{x}(x^{i}-\ _{1}x^{i})^{\alpha }\ _{\ _{1}x}%
\overset{\alpha }{\underline{\partial }}_{i}
\end{equation*}%
when the fractional 1--form (\ref{fr1f}) is$\ \overset{\alpha }{\omega }=%
\frac{1}{\Gamma (\alpha +1)}\ _{\ _{1}x}\overset{\alpha }{d}_{x}(x^{i}-\
_{1}x^{i})^{\alpha }\ F_{i}(x).$

The above formulas introduce a well defined exterior calculus of fractional
differential forms on flat spaces $\mathbb{R}^{n};$ we can generalize the
constructions for a real manifold \ $M,\dim M=n.$\ Any charts of a covering
atlas of $M$ can be endowed with a fractional derivative--integral structure
of Caputo type as we explained above. Such a space $\overset{\alpha }{%
\underline{M}}$ \ (derived for a ''prime'' integer dimensional $M$ of
necessary smooth class)$\ $ \ is called a fractional manifold. $\ $%
Similarly, the concept of tangent bundle can be generalized for fractional
dimensions, using the Caputo fractional derivative. A tangent bundle $TM$
over a manifold $M$ is canonically defined by its local integer differential
structure $\partial _{i}$. A fractional version is induced if instead of $%
\partial _{i}$ we consider the left Caputo derivatives $_{\ _{1}x^{i}}%
\overset{\alpha }{\underline{\partial }}_{i}$ of type (\ref{lfcd}), for
every local coordinate $x^{i}$ on $M.$ A fractional tangent bundle $\overset{%
\alpha }{\underline{T}}M$ \ for $\alpha \in (0,1)$ (the symbol $T$ is
underlined in order to emphasize that we shall associate the approach to a
fractional Caputo derivative). For simplicity, we shall write both for
integer and fractional tangent bundles the local coordinates in the form $%
u^{\beta }=(x^{j},y^{j}).$

Any fractional frame basis $\overset{\alpha }{\underline{e}}_{\beta }=e_{\
\beta }^{\beta ^{\prime }}(u^{\beta })\overset{\alpha }{\underline{\partial }%
}_{\beta ^{\prime }}$ on $\overset{\alpha }{\underline{T}}M$ \ is connected
via a vielbein transform $e_{\ \beta }^{\beta ^{\prime }}(u^{\beta })$ with
a fractional local coordinate basis
\begin{equation}
\overset{\alpha }{\underline{\partial }}_{\beta ^{\prime }}=\left( \overset{%
\alpha }{\underline{\partial }}_{j^{\prime }}=_{\ _{1}x^{j^{\prime }}}%
\overset{\alpha }{\underline{\partial }}_{j^{\prime }},\overset{\alpha }{%
\underline{\partial }}_{b^{\prime }}=_{\ _{1}y^{b^{\prime }}}\overset{\alpha
}{\underline{\partial }}_{b^{\prime }}\right) ,  \label{frlcb}
\end{equation}%
for $j^{\prime }=1,2,...,n$ and $b^{\prime }=n+1,n+2,...,n+n.$ The
corresponding fractional co--bases are $\overset{\alpha }{\underline{e}}^{\
\beta }=e_{\beta ^{\prime }\ }^{\ \beta }(u^{\beta })\overset{\alpha }{d}%
u^{\beta ^{\prime }},$ where the fractional local coordinate co--basis is
written
\begin{equation}
\ _{\ }\overset{\alpha }{d}u^{\beta ^{\prime }}=\left( (dx^{i^{\prime
}})^{\alpha },(dy^{a^{\prime }})^{\alpha }\right) ,  \label{frlccb}
\end{equation}%
with h-- and v--components, correspondingly, $(dx^{i^{\prime }})^{\alpha }$
and $(dy^{a^{\prime }})^{\alpha },$ being of type (\ref{fr1f}).

\section{Fractional Lagrange--Finsler Geometry}

\label{s3} A Lagrange space \cite{kern} $L^{n}=(M,L),$ of integer dimension $%
n,$ is defined by a Lagrange fundamental function $L(x,y),$ i.e. a regular
real function $L:$ $TM\rightarrow \mathbb{R},$ for which the Hessian
\begin{equation*}
_{L}g_{ij}=(1/2)\partial ^{2}L/\partial y^{i}\partial y^{j}
\end{equation*}%
is not degenerate\footnote{%
the construction was used for elaborating a model of geometric mechanics
following methods of Finsler geometry in a number of works on
Lagrange--Finsler geometry and generalizations \cite{ma1}, see our
developments for modern gravity and noncommutative spaces \cite{vrflg,vsgg}}.

We can consider that a Lagrange space $L^{n}$ is a Finsler space $F^{n}$ if
and only if its fundamental function $L$ is positive and two homogeneous
with respect to variables $y^{i},$ i.e. $L=F^{2}.$ For simplicity, we shall
work with Lagrange spaces and their fractional generalizations, considering
the Finsler ones to consist of a more particular, homogeneous, subclass.

\begin{definition}
A (target) fractional Lagrange space $\overset{\alpha }{\underline{L^{n}}}=(%
\overset{\alpha }{\underline{M}},\overset{\alpha }{L})$ of \ fractional
dimension $\alpha \in (0,1),$ for a regular real function $\overset{\alpha }{%
L}:$ $\overset{\alpha }{\underline{T}}M\rightarrow \mathbb{R},$  when the
fractional Hessian is
\begin{equation}
\ _{L\ }\overset{\alpha }{g}_{ij}=\frac{1}{4}\left( \overset{\alpha }{%
\underline{\partial }}_{i}\overset{\alpha }{\underline{\partial }}_{j}+%
\overset{\alpha }{\underline{\partial }}_{j}\overset{\alpha }{\underline{%
\partial }}_{i}\right) \overset{\alpha }{L}\neq 0.  \label{hessf}
\end{equation}
\end{definition}

In our further constructions, we shall use the coefficients $\ _{L\ }\overset%
{\alpha }{g^{ij}\text{ }}$being inverse to $_{L\ }\overset{\alpha }{g}_{ij}$
(\ref{hessf}).\footnote{%
We shall put a left label $L$ to certain geometric objects if it is
necessary to emphasize that they are induced by Lagrange generating
function. Nevertheless, such labels will be omitted (in order to simplify the
notations) if that will not result in ambiguities.} Any $\overset{\alpha }{%
\underline{L^{n}}}$ can be associated to a prime ''integer'' Lagrange space $%
L^{n}.$

The concept of nonlinear connection (N--connection) on $\overset{\alpha }{%
\underline{L^{n}}}$ can be introduced similarly to that on nonholonomic
fractional manifold \cite{vrfrf,vrfrg} considering the fractional tangent
bundle $\overset{\alpha }{\underline{T}}M.$

\begin{definition}
A N--connection $\overset{\alpha }{\mathbf{N}}$ on $\overset{\alpha }{%
\underline{T}}M$ is defined by a nonholonomic distribution (Whitney sum)
with conventional h-- and v--subspaces, $\underline{h}$ $\overset{\alpha }{%
\underline{T}}M$ and $\underline{v}\overset{\alpha }{\underline{T}}M,$ when
\begin{equation}
\ \overset{\alpha }{\underline{T}}\overset{\alpha }{\underline{T}}M=%
\underline{h}\overset{\alpha }{\underline{T}}M\mathbf{\oplus }\underline{v}%
\overset{\alpha }{\underline{T}}M.  \label{whitney}
\end{equation}
\end{definition}

Locally, a fractional N--connection is defined by a set of coefficients, $%
\overset{\alpha }{\mathbf{N}}\mathbf{=}\{\ ^{\alpha }N_{i}^{a}\},$ when
\begin{equation}
\overset{\alpha }{\mathbf{N}}\mathbf{=}\ \ ^{\alpha
}N_{i}^{a}(u)(dx^{i})^{\alpha }\otimes \overset{\alpha }{\underline{\partial
}}_{a},  \label{fnccoef}
\end{equation}%
see local bases (\ref{frlcb}) and (\ref{frlccb}).

By an explicit construction, we prove

\begin{proposition}
There is a class of N--adapted fractional (co) frames linearly depending on $%
\ ^{\alpha }N_{i}^{a},$
\begin{eqnarray}
\ ^{\alpha }\mathbf{e}_{\beta } &=&\left[ \ ^{\alpha }\mathbf{e}_{j}=\overset%
{\alpha }{\underline{\partial }}_{j}-\ ^{\alpha }N_{j}^{a}\overset{\alpha }{%
\underline{\partial }}_{a},\ ^{\alpha }e_{b}=\overset{\alpha }{\underline{%
\partial }}_{b}\right] ,  \label{dder} \\
\ ^{\alpha }\mathbf{e}^{\beta } &=&[\ ^{\alpha }e^{j}=(dx^{j})^{\alpha },\
^{\alpha }\mathbf{e}^{b}=(dy^{b})^{\alpha }+\ ^{\alpha
}N_{k}^{b}(dx^{k})^{\alpha }].  \label{ddif}
\end{eqnarray}
\end{proposition}

The above bases are nonholonomical (equivalently, non--integrable/
anholonomic) and characterized by the property that
\begin{equation*}
\left[ \ ^{\alpha }\mathbf{e}_{\alpha },\ ^{\alpha }\mathbf{e}_{\beta }%
\right] =\ ^{\alpha }\mathbf{e}_{\alpha }\ ^{\alpha }\mathbf{e}_{\beta }-\
^{\alpha }\mathbf{e}_{\beta }\ ^{\alpha }\mathbf{e}_{\alpha }=\ ^{\alpha
}W_{\alpha \beta }^{\gamma }\ ^{\alpha }\mathbf{e}_{\gamma },
\end{equation*}%
where the nontrivial nonholonomy coefficients $\ ^{\alpha }W_{\alpha \beta
}^{\gamma }$ are computed $\ ^{\alpha }W_{ib}^{a}=\overset{\alpha }{%
\underline{\partial }}_{b}\ ^{\alpha }N_{i}^{a}$ and $\ ^{\alpha
}W_{ij}^{a}=\ ^{\alpha }\Omega _{ji}^{a}=\ ^{\alpha }\mathbf{e}_{i}\
^{\alpha }N_{j}^{a}-\ ^{\alpha }\mathbf{e}_{j}\ ^{\alpha }N_{i}^{a};$ the
values $\ ^{\alpha }\Omega _{ji}^{a}$ define the coefficients of the
N--connection curvature.

Let$\ $\ us consider values $y^{k}(\tau )=dx^{k}(\tau )/d\tau ,$ for $x(\tau
)$ parametrizing smooth curves on a manifold $M$ \ with $\tau \in \lbrack
0,1].$ The fractional analogs of such configurations are determined by
changing $\ d/d\tau $ into the fractional Caputo derivative $\ \overset{%
\alpha }{\underline{\partial }}_{\tau }=_{\ _{1}\tau }\overset{\alpha }{%
\underline{\partial }}_{\tau }$when $\ ^{\alpha }y^{k}(\tau )=\overset{%
\alpha }{\underline{\partial }}_{\tau }x^{k}(\tau ).$ For simplicity, we
shall omit the label $\alpha $ for $y\in $ $\overset{\alpha }{\underline{T}}%
M $ if that will not result in ambiguities and/or we shall do not associate
to it an explicit fractional derivative along a curve.

By straightforward computations, following the same schemes as in \cite%
{ma1,vrflg,vsgg} but with fractional derivatives and integrals, we prove:

\begin{theorem}
Any $\overset{\alpha }{L}$ defines the fundamental geometric objects
determining canonically a nonholonomic fractional Riemann--Cartan geometry
on $\overset{\alpha }{\underline{T}}M$ being satisfied the properties:

\begin{enumerate}
\item The fractional Euler--Lagrange equations%
\begin{equation*}
\ \overset{\alpha }{\underline{\partial }}_{\tau \ }(\ _{\ _{1}y^{i}}\overset%
{\alpha }{\underline{\partial }}_{i}\overset{\alpha }{L})-_{\ _{1}x^{i}}%
\overset{\alpha }{\underline{\partial }}_{i}\overset{\alpha }{L}=0
\end{equation*}
are equivalent to the fractional ''nonlinear geodesic'' (equivalently,
semi--spray) equations $\ $%
\begin{equation*}
\left( \overset{\alpha }{\underline{\partial }}_{\tau \ }\right) ^{2}x^{k}+2%
\overset{\alpha }{G^{k}}(x,\ ^{\alpha }y)=0,
\end{equation*}
where
\begin{equation*}
\overset{\alpha }{G^{k}}=\frac{1}{4}\ \ _{L\ }\overset{\alpha }{g^{kj}}\left[
y^{j}\ _{\ _{1}y^{j}}\overset{\alpha }{\underline{\partial }}_{j}\ \left(
_{\ _{1}x^{i}}\overset{\alpha }{\underline{\partial }}_{i}\overset{\alpha }{L%
}\right) -\ _{\ _{1}x^{i}}\overset{\alpha }{\underline{\partial }}_{i}%
\overset{\alpha }{L}\right]
\end{equation*}%
defines the canonical N--connection $\ $%
\begin{equation}
\ _{L}^{\alpha }N_{j}^{a}=\ _{\ _{1}y^{j}}\overset{\alpha }{\underline{%
\partial }}_{j}\overset{\alpha }{G^{k}}(x,\ ^{\alpha }y).  \label{cncl}
\end{equation}

\item There is a canonical (Sasaki type) metric structure,
\begin{equation}
\ \ _{L}\overset{\alpha }{\mathbf{g}}=\ _{L}^{\alpha }g_{kj}(x,y)\ ^{\alpha
}e^{k}\otimes \ ^{\alpha }e^{j}+\ _{L}^{\alpha }g_{cb}(x,y)\ _{L}^{\alpha }%
\mathbf{e}^{c}\otimes \ _{L}^{\alpha }\mathbf{e}^{b},  \label{sasm}
\end{equation}%
where the preferred frame structure (defined linearly by $\ \ _{L}^{\alpha
}N_{j}^{a})$ is $\ _{L}^{\alpha }\mathbf{e}_{\nu }=(\ _{L}^{\alpha }\mathbf{e%
}_{i},e_{a}).$\footnote{%
A (fractional) general metric structure $\overset{\alpha }{\mathbf{g}}=\{\
^{\alpha }g_{\underline{\alpha }\underline{\beta }}\}$ is defined on $%
\overset{\alpha }{\underline{T}}M$ \ by a symmetric second rank tensor $%
\overset{\alpha }{\mathbf{g}}=\ ^{\alpha }g_{\underline{\gamma }\underline{%
\beta }}(u)(du^{\underline{\gamma }})^{\alpha }\otimes (du^{\underline{\beta
}})^{\alpha }.$ For N--adapted constructions, see details in \cite%
{vrfrf,vrfrg}, it is important to use the property that any fractional
metric $\overset{\alpha }{\mathbf{g}}$ can be represented equivalently as a
distinguished metric (d--metric), $\ \overset{\alpha }{\mathbf{g}}=\left[ \
^{\alpha }g_{kj},\ ^{\alpha }g_{cb}\right] ,$ when
\begin{eqnarray*}
\ \overset{\alpha }{\mathbf{g}} &=&\ ^{\alpha }g_{kj}(x,y)\ ^{\alpha
}e^{k}\otimes \ ^{\alpha }e^{j}+\ ^{\alpha }g_{cb}(x,y)\ ^{\alpha }\mathbf{e}%
^{c}\otimes \ ^{\alpha }\mathbf{e}^{b} \\
&=&\eta _{k^{\prime }j^{\prime }}\ ^{\alpha }e^{k^{\prime }}\otimes \
^{\alpha }e^{j^{\prime }}+\eta _{c^{\prime }b^{\prime }}\ ^{\alpha }\mathbf{e%
}^{c^{\prime }}\otimes \ ^{\alpha }\mathbf{e}^{b^{\prime }},
\end{eqnarray*}%
where matrices $\eta _{k^{\prime }j^{\prime }}=diag[\pm 1,\pm 1,...,\pm 1]$
and $\eta _{a^{\prime }b^{\prime }}=diag[\pm 1,\pm 1,...,\pm 1]$ (reflecting
signature of ''prime'' space $TM$) are obtained by frame transforms $\eta
_{k^{\prime }j^{\prime }}=e_{\ k^{\prime }}^{k}\ e_{\ j^{\prime }}^{j}\ _{\
}^{\alpha }g_{kj}$ and $\eta _{a^{\prime }b^{\prime }}=e_{\ a^{\prime
}}^{a}\ e_{\ b^{\prime }}^{b}\ _{\ }^{\alpha }g_{ab}.$ For fractional
computations, it is convenient to work with constants $\eta _{k^{\prime
}j^{\prime }}$ and $\eta _{a^{\prime }b^{\prime }}$ because the Caputo
derivatives of constants are zero. This allows us to keep the same tensor
rules as for the integer dimensions even the rules for taking local
derivatives became more sophisticate because of N--coefficients $\ ^{\alpha
}N_{i}^{a}(u)$ and additional vierbein transforms $e_{\ k^{\prime }}^{k}(u)$
and $e_{\ a^{\prime }}^{a}(u).$ Such coefficients mix fractional derivatives
$\overset{\alpha }{\underline{\partial }}_{a}$ computed as a local
integration (\ref{lfcd}). If we work with RL fractional derivatives, the
computation become very sophisticate with nonlinear mixing of integration,
partial derivatives etc. For some purposes, just RL may be very important.
Our ideas, for such cases, is to derive a geometric model with the Caputo
factional derivatives, and after that we can re--define the nonholonomic
frames/distributions in a form to extract certain constructions with the RL
derivative.}

\item There is a canonical metrical distinguished connection
\begin{equation*}
\ _{c}^{\alpha }\mathbf{D}=(h\ _{c}^{\alpha }D,v\ _{c}^{\alpha }D)=\{\
_{c}^{\alpha }\mathbf{\Gamma }_{\ \alpha \beta }^{\gamma }=(\ ^{\alpha }%
\widehat{L}_{\ jk}^{i},\ ^{\alpha }\widehat{C}_{jc}^{i})\}\ ,
\end{equation*}%
(in brief, d--connection), which is a linear connection preserving under
parallelism the splitting (\ref{whitney}) and metric compatible, i.e. $\
_{c}^{\alpha }\mathbf{D}\ \left( \ \ _{L}\overset{\alpha }{\mathbf{g}}%
\right) =0,$ when
\begin{equation*}
\ _{c}^{\alpha }\mathbf{\Gamma }_{\ j}^{i}=\ _{c}^{\alpha }\mathbf{\Gamma }%
_{\ j\gamma }^{i}\ _{L}^{\alpha }\mathbf{e}^{\gamma }=\widehat{L}_{\
jk}^{i}e^{k}+\widehat{C}_{jc}^{i}\ _{L}^{\alpha }\mathbf{e}^{c},
\end{equation*}%
for $\widehat{L}_{\ jk}^{i}=\widehat{L}_{\ bk}^{a},\widehat{C}_{jc}^{i}=%
\widehat{C}_{bc}^{a}$ in $\ \ _{c}^{\alpha }\mathbf{\Gamma }_{\ b}^{a}=\
_{c}^{\alpha }\mathbf{\Gamma }_{\ b\gamma }^{a}\ _{L}^{\alpha }\mathbf{e}%
^{\gamma }=\widehat{L}_{\ bk}^{a}e^{k}+\widehat{C}_{bc}^{a}\ _{L}^{\alpha }%
\mathbf{e}^{c},$ and
\begin{eqnarray}
\ ^{\alpha }\widehat{L}_{jk}^{i} &=&\frac{1}{2}\ _{L}^{\alpha }g^{ir}\left(
\ _{L}^{\alpha }\mathbf{e}_{k}\ _{L}^{\alpha }g_{jr}+\ _{L}^{\alpha }\mathbf{%
e}_{j}\ _{L}^{\alpha }g_{kr}-\ _{L}^{\alpha }\mathbf{e}_{r}\ _{L}^{\alpha
}g_{jk}\right) ,  \label{cdc} \\
\ \ ^{\alpha }\widehat{C}_{bc}^{a} &=&\frac{1}{2}\ _{L}^{\alpha
}g^{ad}\left( \ ^{\alpha }e_{c}\ _{L}^{\alpha }g_{bd}+\ ^{\alpha }e_{c}\
_{L}^{\alpha }g_{cd}-\ ^{\alpha }e_{d}\ _{L}^{\alpha }g_{bc}\right)  \notag
\end{eqnarray}%
are just the generalized Christoffel indices.\footnote{%
for integer dimensions, we contract ''horizontal'' and ''vertical'' indices
following the rule: $i=1$ is $a=n+1;$ $i=2$ is $a=n+2;$ ... $i=n$ is $a=n+n"$%
}
\end{enumerate}
\end{theorem}

Finally, in this section, we note that:

\begin{remark}
We note that $\ _{c}^{\alpha }\mathbf{D}$ is with nonholonomically induced
torsion structure defined by 2--forms%
\begin{eqnarray}
\ _{L}^{\alpha }\mathcal{T}^{i} &=&\widehat{C}_{\ jc}^{i}\ ^{\alpha
}e^{i}\wedge \ _{L}^{\alpha }\mathbf{e}^{c},  \label{nztors} \\
\ _{L}^{\alpha }\mathcal{T}^{a} &=&-\frac{1}{2}\ _{L}\Omega _{ij}^{a}\
^{\alpha }e^{i}\wedge \ ^{\alpha }e^{j}+\left( \ ^{\alpha }e_{b}\
_{L}^{\alpha }N_{i}^{a}-\ ^{\alpha }\widehat{L}_{\ bi}^{a}\right) \ ^{\alpha
}e^{i}\wedge \ _{L}^{\alpha }\mathbf{e}^{b}  \notag
\end{eqnarray}%
computed from the fractional version of Cartan's structure equations%
\begin{eqnarray}
d\ ^{\alpha }e^{i}-\ ^{\alpha }e^{k}\wedge \ \ _{c}^{\alpha }\mathbf{\Gamma }%
_{\ k}^{i} &=&-\ _{L}^{\alpha }\mathcal{T}^{i},\   \notag \\
d\ _{L}^{\alpha }\mathbf{e}^{a}-\ _{L}^{\alpha }\mathbf{e}^{b}\wedge \ \
_{c}^{\alpha }\mathbf{\Gamma }_{\ b}^{a} &=&-\ _{L}^{\alpha }\mathcal{T}^{a},
\notag \\
d\ \ _{c}^{\alpha }\mathbf{\Gamma }_{\ j}^{i}-\ \ _{c}^{\alpha }\mathbf{%
\Gamma }_{\ j}^{k}\wedge \ \ _{c}^{\alpha }\mathbf{\Gamma }_{\ k}^{i} &=&-\
_{L}^{\alpha }\mathcal{R}_{j}^{i}  \label{seq}
\end{eqnarray}%
in which the curvature 2--form is denoted $\ _{L}^{\alpha }\mathcal{R}%
_{j}^{i}.$
\end{remark}

In general, for any d--connection on $\ \overset{\alpha }{\underline{T}}M,$
we can compute respectively the N--adapted coefficients of torsion $\
^{\alpha }\mathcal{T}^{\tau }=\{\ ^{\alpha }\mathbf{\Gamma }_{\ \beta \gamma
}^{\tau }\}$ and curvature $\ ^{\alpha }\mathcal{R}_{~\beta }^{\tau }=\{\
^{\alpha }\mathbf{R}_{\ \beta \gamma \delta }^{\tau }\}$ as it is explained
for general fractional nonholonomic manifolds in \cite{vrfrf,vrfrg}.

\section{Almost K\"{a}hler Models of Fractional Lagrange Spaces}

\label{s4}

The goals of this section is to prove that the canonical N--connection $\
_{L}^{\alpha }N$ (\ref{cncl}) induces an almost K\"{a}hler structure defined
canonically by a fractional regular $\overset{\alpha }{L}(x,\ ^{\alpha }y).$

\begin{definition}
A fractional nonholonomic almost complex structure is defined as a linear
operator $\overset{\alpha }{\mathbf{J}}$ acting on the vectors on $\overset{%
\alpha }{\underline{T}}M$ following formulas
\begin{equation*}
\overset{\alpha }{\mathbf{J}}(\ _{L}^{\alpha }\mathbf{e}_{i})=-\ ^{\alpha
}e_{i}\mbox{\ and \ }\overset{\alpha }{\mathbf{J}}(\ ^{\alpha }e_{i})=\
_{L}^{\alpha }\mathbf{e}_{i},
\end{equation*}%
where the superposition $\overset{\alpha }{\mathbf{J}}\mathbf{\circ \overset{%
\alpha }{\mathbf{J}}=-I,}$ for $\mathbf{I}$ being the unity matrix.
\end{definition}

The fractional operator $\overset{\alpha }{\mathbf{J}}$ reduces to a complex
structure $\mathbb{J}$ if and only if the distribution (\ref{whitney}) is
integrable and the dimensions are taken to be some integer ones.

\begin{lemma}
\textbf{--Definition.} A fractional $\overset{\alpha }{L}$ induces a
canonical 1--form $\ \ _{L}^{\alpha }\omega =\frac{1}{2}\left( \ _{\
_{1}y^{i}}\overset{\alpha }{\underline{\partial }}_{i}\overset{\alpha }{L}%
\right) \ ^{\alpha }e^{i}$ and a metric $\ _{L}\overset{\alpha }{\mathbf{g}}$
(\ref{sasm}) induces a canonical 2--form
\begin{equation}
\ \ \ _{L}^{\alpha }\mathbf{\theta }=\ _{L\ }\overset{\alpha }{g}_{ij}(x,\
^{\alpha }y)\ \ _{L}^{\alpha }\mathbf{e}^{i}\wedge \ ^{\alpha }e^{j}.
\label{asstr}
\end{equation}%
associated to $\overset{\alpha }{\mathbf{J}}$ following formulas $\ \ \ \
_{L}^{\alpha }\mathbf{\theta (X,Y)}\doteqdot \ \ \ _{L}\overset{\alpha }{%
\mathbf{g}}\left( \overset{\alpha }{\mathbf{J}}\mathbf{X,Y}\right) $ for any
vectors $\mathbf{X}$ and $\mathbf{Y}$ on $\overset{\alpha }{\underline{T}}M.$
\end{lemma}

Using a fractional N--adapted form calculus,
\begin{eqnarray*}
d\ \ _{L}^{\alpha }\mathbf{\theta } &=&\frac{1}{6}\sum\limits_{(ijk)}\ \
_{L\ }\overset{\alpha }{g}_{is}\ _{L}^{\alpha }\Omega _{jk}^{s}\ ^{\alpha
}e^{i}\wedge \ ^{\alpha }e^{j}\wedge \ ^{\alpha }e^{k} \\
&&+\frac{1}{2}\left( \ \ _{L\ }\overset{\alpha }{g}_{ij\parallel k}-\ \ _{L\
}\overset{\alpha }{g}_{ik\parallel j}\right) \ _{L}^{\alpha }\mathbf{e}%
^{i}\wedge \ ^{\alpha }e^{j}\wedge \ ^{\alpha }e^{k} \\
&&+\frac{1}{2}\left( \ ^{\alpha }e_{k}(\ \ _{L\ }\overset{\alpha }{g}%
_{ij})-\ ^{\alpha }e_{i}\ (\ _{L\ }\overset{\alpha }{g}_{kj})\right) \
_{L}^{\alpha }\mathbf{e}^{k}\wedge \ _{L}^{\alpha }\mathbf{e}^{i}\wedge \
^{\alpha }e^{j},
\end{eqnarray*}%
where $(ijk)$ means symmetrization of indices and
\begin{equation*}
\ \ _{L\ }\overset{\alpha }{g}_{ij\shortparallel k}:=\ _{L}^{\alpha }\mathbf{%
e}_{k}\ \ \ _{L\ }\overset{\alpha }{g}_{ij}-\ ^{L}B_{ik}^{s}\ \ \ _{L\ }%
\overset{\alpha }{g}_{sj}-\ ^{L}B_{jk}^{s}\ \ \ _{L\ }\overset{\alpha }{g}%
_{is},
\end{equation*}%
for $\ ^{L}B_{ik}^{s}=\ ^{\alpha }e_{i}\ \ _{L}^{\alpha }N_{k}^{s},$ we
prove the results:

\begin{proposition}
\begin{enumerate}
\item A regular $L$ defines on $TM$ an almost Hermitian (symplectic)
structure $\ _{L}^{\alpha }\mathbf{\theta }$ for which $d\ \ _{L}^{\alpha
}\omega =\ _{L}^{\alpha }\mathbf{\theta };$

\item The canonical N--connection $\ \ _{L}^{\alpha }N_{j}^{a}$ (\ref{cncl})
and its curvature have the properties
\begin{eqnarray*}
\sum\limits_{ijk}\ \ _{L\ }\overset{\alpha }{g}_{l(i}\ \ _{L}^{\alpha
}\Omega _{jk)}^{l} &=&0,\ \ \ \ _{L\ }\overset{\alpha }{g}_{ij\shortparallel
k}-\ \ _{L\ }\overset{\alpha }{g}_{ik\shortparallel j}=0, \\
\ \ ^{\alpha }e_{k}\ (\ _{L\ }\overset{\alpha }{g}_{ij})-\ ^{\alpha }e_{j}\
(\ _{L\ }\overset{\alpha }{g}_{ik}) &=&0.
\end{eqnarray*}
\end{enumerate}
\end{proposition}

\begin{conclusion}
The above properties state an almost Hermitian model of \ fractional
Lagrange space $\overset{\alpha }{\underline{L^{n}}}=(\overset{\alpha }{%
\underline{M}},\overset{\alpha }{L})$ as a fractional almost K\"{a}hler
manifold with $d\ \ _{L}^{\alpha }\mathbf{\theta }=0.$The triad $\ ^{\alpha }%
\mathbb{K}^{2n}=(\widetilde{\overset{\alpha }{\underline{T}}M},\ \ \ _{L}%
\overset{\alpha }{\mathbf{g}}\mathbf{,}\overset{\alpha }{\mathbf{J}}),$
where $\widetilde{\overset{\alpha }{\underline{T}}M}:=\overset{\alpha }{%
\underline{T}}M\backslash \{0\}$ for $\{0\}$ denoting the null--sections
under $\overset{\alpha }{\underline{M}},$ defines a fractional, and
nonholonomic, almost K\"{a}hler space.
\end{conclusion}

Our next purpose is to construct a canonical (i.e. uniquely determined by $%
\overset{\alpha }{L})$ almost K\"{a}hler distinguished connection
(d--connection) $\ ^{\theta }\overset{\alpha }{\mathbf{D}}$ \ being
compatible both with the almost K\"{a}hler $\left( \ _{L}^{\alpha }\mathbf{%
\theta ,}\overset{\alpha }{\mathbf{J}}\right) $ and N--connection structures
$\ \ _{L}^{\alpha }\mathbf{N}$ when%
\begin{equation}
\ \ ^{\theta }\overset{\alpha }{\mathbf{D}}_{\mathbf{X}}\ \ _{L}\overset{%
\alpha }{\mathbf{g}}\mathbf{=0}\mbox{\ and \ }\mathbf{\ }\ \ ^{\theta }%
\overset{\alpha }{\mathbf{D}}_{\mathbf{X}}\overset{\alpha }{\mathbf{J}}%
\mathbf{=0,}  \label{compcond}
\end{equation}%
for any vector $\mathbf{X}=X^{i}\ \ _{L}^{\alpha }\mathbf{e}_{i}+X^{a}\
^{\alpha }e_{a}.$ By a straightforward computation, we prove (see similar
''integer'' details in \cite{ma1,vrflg,vfedq1}):

\begin{theorem}
\textbf{(Main Result)}\label{th1} The fractional canonical metrical
d--connec\-ti\-on $\ _{c}^{\alpha }\mathbf{D,}$\textbf{\ }possessing
N--adapted coefficients $\left( \widehat{L}_{\ bk}^{a},\widehat{C}%
_{bc}^{a}\right) $ (\ref{cdc}), defines a (uni\-que) canonical fractional
almost K\"{a}hler d--connection $\ _{c}^{\theta }\overset{\alpha }{\mathbf{D}%
}=$ $\ _{c}^{\alpha }\mathbf{D}$ satisfying the conditions (\ref{compcond}).
\end{theorem}

Finally, we note that:

\begin{remark}
There are two important particular cases:

\begin{enumerate}
\item If $\overset{\alpha }{L}=(\overset{\alpha }{F})^{2},$ for a Finsler
space of an integer dimension, we get a canonical almost K\"{a}hler model of
Finsler space \cite{matsumoto}, when $\ \ _{c}^{\theta }\overset{\alpha }{%
\mathbf{D}}=$ $\ _{c}^{\alpha }\mathbf{D}$ transforms in the so--called
Cartan--Finsler connection \cite{cartan} (for integer dimensional Lagrange
spaces, a similar result was proven in \cite{opr1}, see details in \cite{ma1}%
).

\item We get a K\"{a}hlerian model of a Lagrange, or Finsler, space if the
respective almost complex structure $\mathbf{J}$ is integrable. This
property holds also for respective spaces of fractional dimension.
\end{enumerate}
\end{remark}
Finally, we emphasize that the  geometric background for fractional Lagrange--Finsler spaces provided in this section  has been used in our partner work \cite{vdefqfrt} for deformation quantization of such theories. This is an explicit proof that fractional physical models can be quantized in general form,  that it is  possible to elaborate  new types/models/theories of quantum fractional calculus and geometries stating a new unified formalism to fractional classical and quantum geometries and physics and propose various  applications in mechanics and engineering.

\section{Conclusions}

\label{s5} The fractional Caputo derivative allows us to formulate a
self--consistent nonholonomic geometric approach to fractional calculus \cite%
{vrfrf,vrfrg} when various important mathematical and physical problems for spaces and processes of non--integer dimension can be investigated using various methods originally elaborated in modern Lagrange--Hamilton--Finsler geometry and generalizations. In this paper, we provided a geometrization
of regular fractional Lagrange mechanics in a form similar to that in Finsler geometry but, in our case, extended to non--integer dimensions. This way, applications of fractional calculus become a direction of non--Riemannian geometry with   nonholonomic distributions and generalized connections.

Nonholonomic fractional mechanical interactions are described by a sophisticate variational calculus and derived cumbersome systems of nonlinear equations and less defined types of symmetries. To quantize such mechanical systems is a very difficult problem related to a number of unsolved
fundamental problems of nonlinear functional analysis. Nevertheless, for certain approaches based on fractional Caputo derivative, or those admitting nonholonomic deformations/ transforms to such fractional configurations, it
is possible to provide a geometrization of fractional Lagrange mechanics. In this case, the problem of quantization can be approached following powerful methods of geometric and/or deformation quantization.

Having a similarity between fractional Lagrange geometry and certain
generalized almost K\"{a}hler--Finsler models, for which the Fedosov
quantization was performed in our previous works \cite{vfedq1,vfedq2,vfedq3}%
, our main goal was to prove the Theorem \ref{th1}. As a result, we obtained
a suitable geometric background for Fedosov quantization of fractional
Lagrange--Finsler spaces which will be performed in our partner work \cite%
{vdefqfrt}.

Finally, we conclude that the examples of fractional almost K\"{a}hler
spaces considered in this works can be re--defined for another types of
nonholonomic distributions and applied, for instance, as a geometric scheme
for deformation/ A--brane quantization of fractional Hamilton and Einstein
spaces and various generalizations, similarly to our former constructions
performed for integer dimensions in Refs. \cite{vfed4,anastv,vbrane}. Such
developments consist a purpose of our further research.

\vskip5pt \textbf{Acknowledgement: } S. V. is grateful to \c{C}ankaya
University for support of his research on fractional calculus, geometry and
applications.

\end{document}